\documentclass[a4paper]{spie}  
\usepackage[]{graphicx}

\title{The Gaia payload uplink commanding system } 

\author{
A.~Mora\supit{a}\supit{b},
A.~Abreu\supit{a}\supit{c},
N.~Cheek\supit{a}\supit{d},
C.M.~Crowley\supit{a}\supit{e},
R.~Guerra\supit{a}\supit{f},
J.~Hern\'andez\supit{a},
E.~Joliet\supit{a}\supit{e}.
R.~Kohley\supit{a},
J.~Mart\'{\i}n-Fleitas\supit{a}\supit{b},
J.~Osinde\supit{a}\supit{g} and
H.~Siddiqui\supit{a}\supit{h}
\skiplinehalf
\supit{a}ESA-ESAC Gaia SOC, P.O. Box 78, 28691 Villanueva de la Ca\~{n}ada, Madrid, Spain; \\
\supit{b}Aurora Technology, Crown Business Centre, Heereweg 345, 2161 CA Lisse, The Netherlands; \\
\supit{c}Elecnor Deimos Space. Ronda de Poniente 19. Edificaci\'on Fiteni VI. 28760 Tres Cantos, Madrid, Spain; \\
\supit{d}Serco Gesti\'on de Negocios. Valle del Roncal 12, 28232 Las Rozas, Madrid, Spain; \\
\supit{e}HE Space Operations BV, Huygensstraat 44, 2201 DK Noordwijk, The Netherlands; \\
\supit{f}GMV C/ Isaac Newton, 11, 28760 Tres Cantos, Madrid, Spain; \\
\supit{g}Isdefe, Calle de Beatriz de Bobadilla 3, 28040 Madrid, Spain; \\
\supit{h}Telespazio VEGA UK Ltd, 350 Capability Green, Luton, Bedfordshire, LU1 3LU, UK; \\
}

\authorinfo{Further author information: (Send correspondence to A.M.)\\A.M.: E-mail: alcione.mora@esa.int, Telephone: +34 91 813 1480}

\begin{document} 
  \maketitle 

\begin{abstract}
This document describes the uplink commanding system for the ESA Gaia mission. The need for commanding, the main actors, data flow and systems involved are described. The system architecture is explained in detail, including the different levels of configuration control, software systems and data models. A particular subsystem, the automatic interpreter of human-readable onboard activity templates, is also carefully described. Many lessons have been learned during the commissioning and are also reported, because they could be useful for future space survey missions.
\end{abstract}




\section{Introduction}

Gaia is a mission of the European Space Agency (ESA) carrying out a global astrometric survey of the Galaxy. It was launched on 19 December 2013 and is currently finishing commissioning\cite{2014SPIE.Prusti}\cite{2014SPIE.Els}. One key difference with respect to other missions is the absence of principal investigators. ESA has procured the whole spacecraft, including the instruments (focal planes) to industry. The main contractor is Airbus Defence \& Space, Toulouse.

The survey strategy is fully defined by the scanning law followed by the spacecraft to observe the sky in Time Delayed Integration (TDI) mode with its gigapixel CCD mosaic camera. In this way, there are no schedulers, as in typical observatory missions, where continuous human decisions are required to produce an efficient observing strategy. However, payload commanding is still required for Gaia regarding four different aspects.

First, Gaia does not have an input catalogue. The decisions of what objects to observe, and how to distribute the available resources, are taken in real time by a set of seven on-board computers, the Video Processing Units (VPUs). Those computers are continuously running complex algorithms in charge of e.g. object detection and confirmation, magnitude estimation, CCD read-out strategy, telemetry packet generation, etc.... The behaviour of each VPU is highly customizable via a 725.7 kB parameter block divided into 150 tables.

Second, there is a way to force Gaia do some custom fake observations using the so-called Virtual Objects (VOs). They are very useful to e.g. determine the sky background and radiation damage when combined with CCD charge injections. The number, frequency and properties of VOs are controlled by a particular VPU table. There is an on-board storage area for a number of pre-loaded VO patterns. Currently, there are 24 such schemes per VPU, with a size of 64 kB each.

Third, all the payload telemetry is stored in the Payload Data Handling Unit (PDHU) before downlink. This system stores data in files. Each file has different properties, such as maximum size, preallocated circular buffer or dynamic storage, downlink and deletion priorities, etc.... The PDHU behaviour is controlled by a number of small binary tables, which can be modified according to the mission evolution.

Fourth, the observer Service Interface Function (SIF) records portions of the VPU memory for subsequent downlink and analysis. The SIF was originally envisioned as a pure engineering troubleshooting mode. However, some DPAC operations rely on the SIF to provide useful data not routinely downlinked, such as post-scan pixels.

\begin{figure}
\includegraphics[width=\hsize]{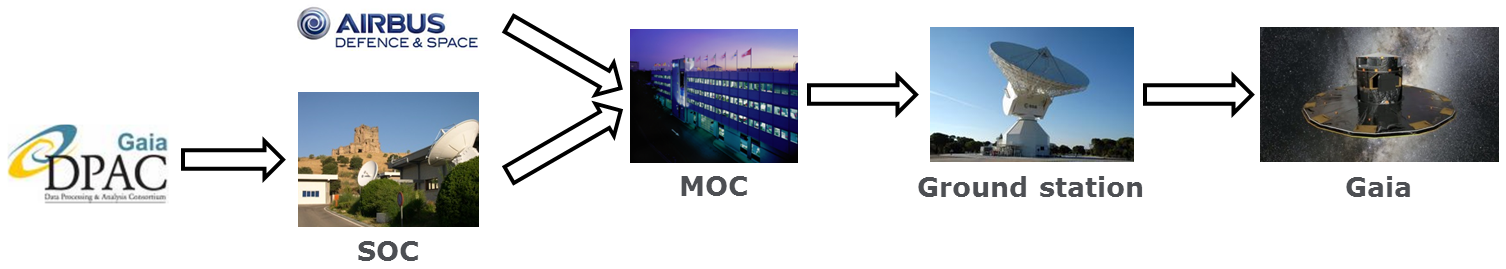}
\caption{Uplink data flow. Uplink requests are either proposed by the Gaia Data processing and Analysis Consortium (DPAC) and handled by the Science Operations Centre (SOC) or directly generated by  Airbus Defence \& Space (prime contractor). Both inputs are transmitted to the Mission Operations Centre (MOC), which validates and converts them into telecommands, which are subsequently uplinked to Gaia via the ESA deep space network of ground stations.}
\label{fig:uplinkDataFlow}
\end{figure}

The total size of the VPU parameters, VO patterns and PDHU tables is small (around 16 MB). However, they are critical both to define the spacecraft behaviour and to correctly interpret the telemetry by the downlink processing pipeline (see\cite{2014SPIE.Siddiqui} for a detailed description of the downlink systems). Very well defined interfaces have been developed to command changes onboard, put them under configuration control and subsequently distribute them to the data processing centres and members of the Gaia Data Processing and Analysis Consortium\cite{LL:FM-030}, \cite{2008IAUS..248..224M}\cite{2007ASPC..376...99O} (DPAC). An schematic overview of the uplink process is provided in Fig.~\ref{fig:uplinkDataFlow}.

In this document, a detailed overview on the uplink system architecture will be provided in Sect.~\ref{sect:workFlow}. Additional information will be provided in Sect.~\ref{sect:scpLanguageParser} on the automatic system put in place to handle complex payload operations in a human readable format. Some problems and lessons learned during the commissioning are also be presented in Sect.~\ref{sect:lessonsLearned}. Finally, the conclusions are provided in Sect.~\ref{sect:conclusions}.

\section{Work flow description}
\label{sect:workFlow}

DPAC Payload Operations Requests (PORs) are handled by the Science Operations Centre\cite{2008IAUS..248..282O} (SOC)  Payload Operations System (POS). POS and all software packages described in this proceeding have been developed in Java. Object data bases are used to exchange and store information. Complex data models are used to define the classes and can be edited and visualized using the Main Data Base Dictionary Tool. In this document, many screen captures from the Dictionary Tool will be used to better explain the structure of the main classes used.

\begin{figure}
\includegraphics[trim=1.5cm 2.5cm 1.5cm 2.8cm, clip=true,width=\hsize]{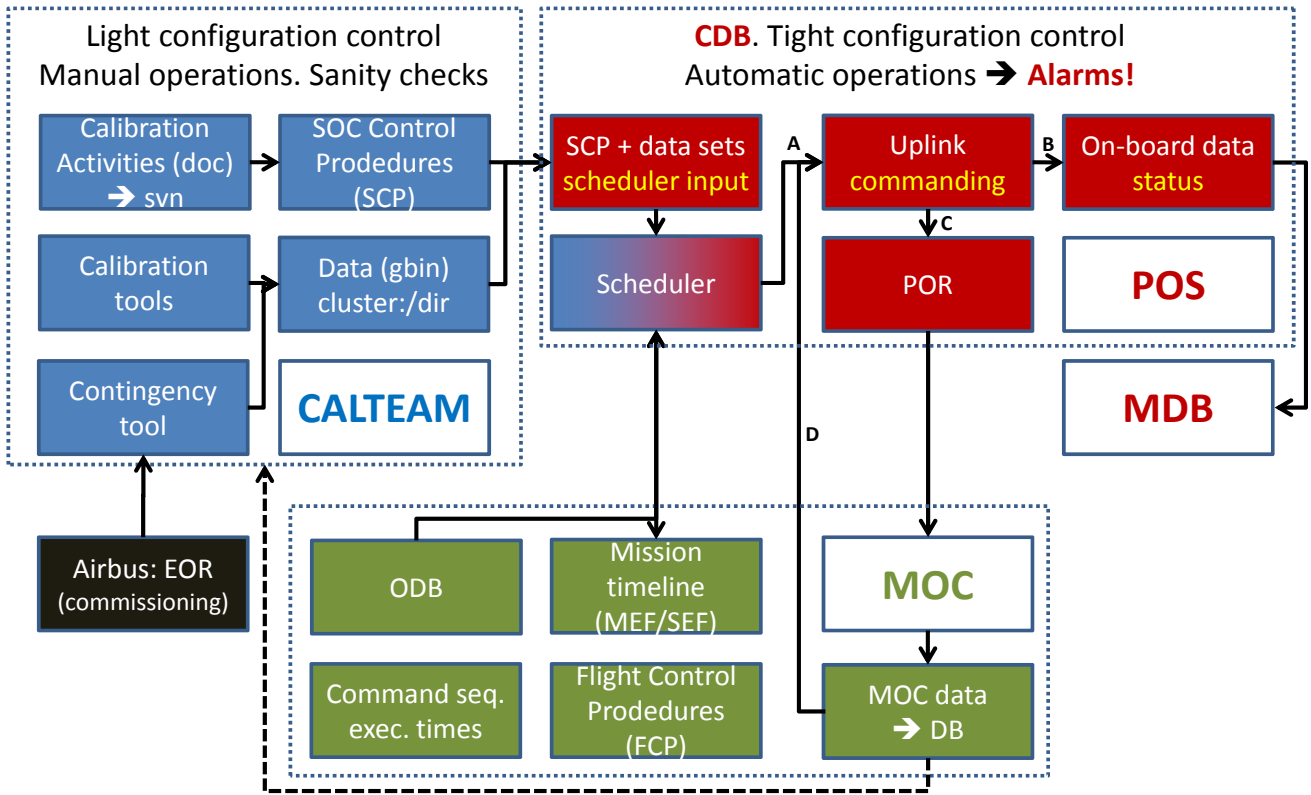}
\caption{On-board data uplink process overview. The key element is the scheduler, which drives the generation of new tables, data ingestion in the Configuration Data Base (CDB) and Payload Operations Request (POR) generation. The most important Payload Operations System (POS) activities are identified with letters A-D. A: The scheduler prepares the uplink using SCPs, new data and MOC constraints. B: Future on-board data are stored in the CDB. C: PORs are generated for each uplink using CDB data. D: MOC data validate the uplink and change status flags. Problems are solved using new ``contingency'' uplink and CDB objects. Invalid uplink and on-board data are flagged and UTC time tagged, but not retracted from the CDB. Finally, Airbus commanding during commissioning is also tracked in the CDB to correctly process the data by DPAC.}
\label{fig:functionalDiagram}
\end{figure}

Fig.~\ref{fig:functionalDiagram} provides an overview of the whole process. The top blue and red boxes enclosed in a dotted line describe the DPAC elements directly involved. The blue area is the SOC calibration team parser-ingestor, and is under light configuration control. The red area is part of the ESAC Data Processing Centre (DPCE) POS system, and is under tight configuration control. All data handled by the POS is stored in the Configuration Data Base (CDB). The green boxes represent elements provided by MOC. The MDB red box outside the dotted areas describes the interaction with other DPAC systems via the MDB.

The key element in the whole system are: the {\bf data}, {\bf SCP} and {\bf Scheduler}. The data are the pieces of information to be updated onboard and to be put under configuration control. They are updated using SCPs (SOC Control Procedures), predefined templates including a series of telecommand sequences. Finally, the scheduler is the high level interface with POS. It starts the process of commanding the VPU, generating new VO patterns, requesting SIF or updating on-board data using the input data and SCPs.

\begin{figure}
\begin{center}
\includegraphics[scale=0.45]{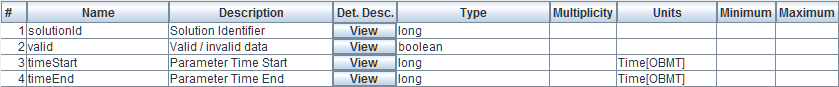}
\end{center}
\caption{Data model for the {\tt OnboardData} class. This is the parent class for {\tt VpuParameter}, {\tt PdhuTable} and {\tt VoPatternStore}. It includes a validity time range and flag. The latter indicates whether the datum really represents something happening onboard or a past misconception by the ground segment.}
\label{fig:onboardData}
\end{figure}

The data model developed to describe the on-board data, VPU tables, PDHU tables and VO patterns storage is composed of the classes {\tt OnBoardData}. {\tt VpuParameter}, {\tt PdhuTable} and {\tt VoPatternStore}, respectively. The latter three classes extend {\tt OnBoardData}. See Fig.~\ref{fig:onboardData}. {\tt Vpu\-Parameter} and {\tt PdhuTable} are also extended by different subclasses. {\tt SifCommand} provides a logical representation of the SIF commands sent to the VPUs. Whenever a full or partial change is applied to an on-board datum, a whole new object will be produced and stored. That is, {\tt OnBoardData} represents the quantum of information used to specify any change in the spacecraft.

\begin{figure}
\begin{center}
\includegraphics[scale=0.45]{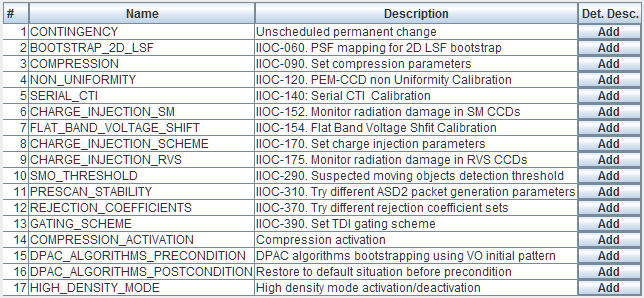}
\end{center}
\caption{Examples of calibration activities used during commissioning. They are defined in DPAC technical notes. Each calibration activity is implemented as a different SCP for each VPU.}
\label{fig:calibrationActivity}
\end{figure}

\begin{figure}
\begin{center}
\includegraphics[scale=0.45]{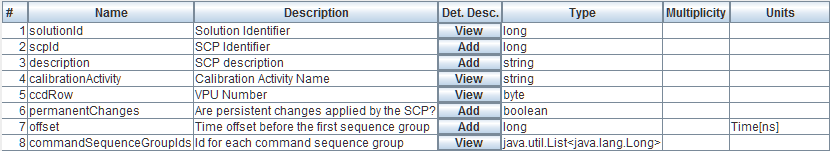}
\end{center}
\caption{Data model for the {\tt Scp} class. It represents the SOC Calibration Procedures (SCPs). Each SCP is uniquely identified by the calibration activity and CCD row (VPU). They are templates describing a given calibration activity for a given spacecraft status and VPU (if applicable). Each SCP is composed of several {\tt CommandSequenceGroup}. A waiting offset time may be applied at the beginning. The description field is used by the POS to construct the POR header. Different versions of a given SCP are created each time the parser-ingestor is run, and are distinguished by the solution ID. An unique numerical scpId is assigned to each SCP and is preserved along versions. Old invalid or deprecated SCPs are kept in the CDB, but the parser-ingestor does not regenerate them (and assign new solution IDs) in subsequent executions.}
\label{fig:scp}
\end{figure}

Each DPAC onboard activity is represented by a {\bf Calibration activity}, the high level documents specifying them, Some examples are listed in Fig.~\ref{fig:calibrationActivity}. The detailed logical templates defining the different steps implementing a calibration activity are the {\bf SOC Control Procedures (SCPs)}. The scheduler fills each SCP template using the data sets provided by the calibration tools. SCPs have the structure shown in Fig.~\ref{fig:scp}. Changes not commanded by the SOC (Airbus, MOC, problems) are handled by the {\bf Contingency tool}, whose output is stored. No contingency tool keeps track of SIF commands not originated from DPAC.

\begin{figure}
\begin{center}
\includegraphics[scale=0.45]{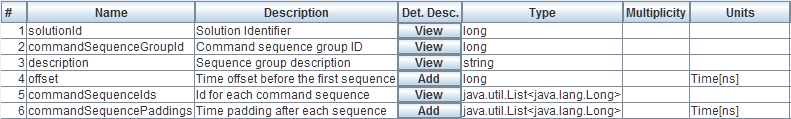}
\end{center}
\caption{Data model for the {\tt CommandSequenceGroup} class. It represents a set of related and consecutive command sequences. A waiting offset time may be applied at the beginning. Padding waiting times can also be used after each sequence. The description text is also unique. Invalid groups are kept for future reference, although not regenerated by the parser-ingestor any more. The unique command sequence group ID is preserved over all versions, controlled by the solution ID.}
\label{fig:commandSequenceGroup}
\end{figure}

Each SCP is uniquely identified by the calibration activity and CCD row (VPU). Different versions of a given SCP are created each time the parser-ingestor is run, and are distinguished by the solution ID. An unique numerical scpId is assigned to each SCP and is preserved along versions. Old invalid or deprecated SCPs are kept in the CDB, but the parser-ingestor does not regenerate them (and assign new solution IDs) in subsequent executions. A text description is used to construct the POR header information. Even though several VPUs can run some calibration activities, each VPU has its own SCP, in agreement with MOC procedures. The core of the SCP is a list of {\tt Command\-Sequence\-Group} IDs, see Fig.~\ref{fig:commandSequenceGroup}. Each of these objects represents a group of related consecutive command sequences. A waiting offset time may be applied at the beginning. A flag indicates whether permanent on-board changes are expected after the SCP. It is used by the POS to check the CDB consistency. Each {\tt Command\-Sequence\-Group} is also composed of a list of command sequence identifiers plus leading offset and padding waiting times, defined in the same way as the SCPs.

\begin{figure}
\begin{center}
\includegraphics[scale=0.45]{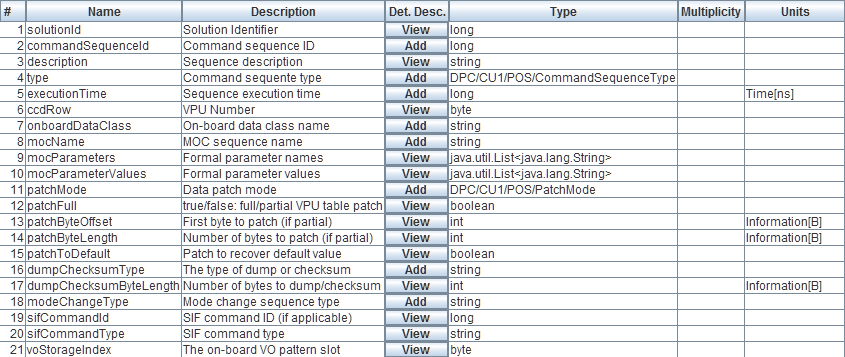}
\end{center}
\caption{Data model for the {\tt CommandSequence} class. It includes the MOC command sequence name from the Onboard Data Base (ODB), a textual description, the execution time, the type of data to update/dump/checksum (if applicable) the type of data update to make (full or partial), and the portion of data to patch, if partial.}
\label{fig:commandSequence}
\end{figure}

\begin{figure}
\begin{center}
\includegraphics[scale=0.45]{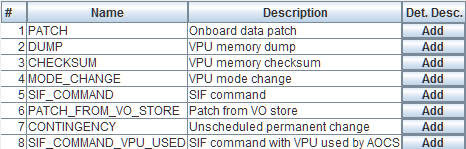}
\end{center}
\caption{Data model for the {\tt CommandSequenceType} class. It is an enumeration describing the purpose of the command sequence.}
\label{fig:commandSequenceType}
\end{figure}

The command sequences themselves are described by objects of the class {\tt Command\-Sequence} (see Fig.~\ref{fig:commandSequence}). They are composed of several fields. The unique ID is used by the {\tt Command\-Sequence\-Group} and is preserved over all versions, controlled by the solution ID. A description field is used to provide additional information inside the POR file. The type is an enumeration describing if it commands the change (patch) of some on-board data (most command sequences are of this type), a spacecraft mode change such as set VPU to service/operational mode or include/exclude VPU from Attitude and Orbit Control System (AOCS) loop, VPU table dump, VPU table checksum, SIF command, VO VPU table patch using the on-board VO pattern store or accounts for contingency non-SOC scheduled on-board changes (see Fig.~\ref{fig:commandSequenceType}). Note that this is a simplified abstraction of true command sequences, which are concatenations of spacecraft telecommands handled by MOC and provided with a number of tunable (formal) parameters.

Most on-board data changes involve the full update of a VPU or PDHU table or a VO pattern in the on-board storage. However, this is sometimes inconvenient. In particular, VPU table \#78 (hardware parameters) is very large and MOC has implemented command sequences for full and partial updates. Information on the type of update (full or partial), the patch mode (BDT, short load) and the portion of bytes to patch is also included. The mode used to update a full VPU table is called Bulk Data Transfer (BDT), as opposed to short logical load, which patches small areas of the memory. This allows e.g. to simultaneously partially patch several VPU tables. However, DPAC uses BDT when possible. The field patchToDefault indicates whether a patch is supposed to restore the configuration before SCP execution.

\begin{figure}
\begin{center}
\includegraphics[scale=0.45]{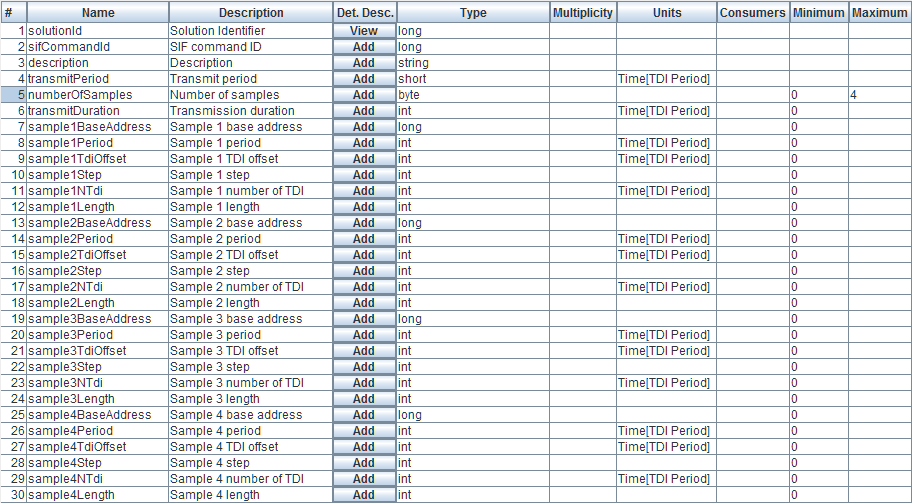}
\end{center}
\caption{Data model for the {\tt SifCommand} class. It is a logical representation of a SIF command.}
\label{fig:sifCommand}
\end{figure}

Only a limited set of pre-defined dump, checksum, mode change and SIF commands are implemented for DPAC. Mode changes, dumps and checksums are identified by a string, while immutable SIF commands are identified by the unique SIF command ID. {\tt SifCommand} object (see Fig.~\ref{fig:sifCommand}). This limited choice makes constructing and debugging SCPs much easier.

All the previous fields define the functionality of a command sequence. There can not be two different command sequences implementing the same functionality with the same solution ID. The functionality of a command sequence is implemented by its structure, which for SOC purposes is composed of the MOC name, formal parameter names and values. The structure is described in detail in the {\bf Operational Data Base, ODB}. The formal parameter values are stored as ASCII values and unambiguously interpreted by the POS according to the ODB data type. For all non-patch commands, the parameters to provide are stored in the data base as a list of immutable values. That is, the formal parameters of the underlying MOC command sequences are frozen. The parameters required by a given command sequence may evolve during the mission. Those changes will be reflected into the ODB. The strategy adopted to ensure compatibility with the ODB is using it, together with additional information to automatically produce the {\tt CommandSequence} objects using ad-hoc deterministic code integrated within the parser-ingestor.

When new SCPs are created, example PORs are sent to MOC for validation. Invalid SCPs are flagged accordingly and stored for future reference. The {\bf Command sequence execution times} are estimated using MOC guidelines, which include overheads over the basic execution time to avoid the impossible simultaneous execution of more than one telecommand. The scheduler has certain windows available in the {\bf Mission timeline} to program SCP executions. The ODB, mission timeline and command execution times are inputs provided by MOC.

\begin{figure}
\begin{center}
\includegraphics[scale=0.45]{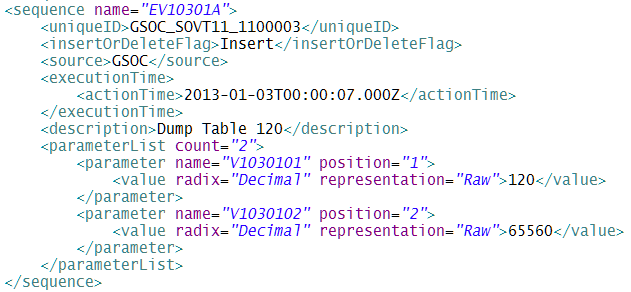}
\end{center}
\caption{Excerpt from a POR. It is an xml file providing the times when several MOC command sequences should be executed. The parameters for each command are provided in the POR. To avoid errors, the parameter input values will be pre-stored for each command sequence, except on-board data patches. Some description attributes and fields are allowed within the POR to improve legibility. They are used as proxies for a browser xsl style sheet.}
\label{fig:porSample}
\end{figure}

For each executed SCP, the scheduler carries out three actions. First, any new on-board data values are stored in the {\bf CDB} as new {\tt OnBoardData} objects. Each one belongs to a CDB data set, previously generated by a calibration tool. Validity times in the future consistent with the POR are assigned, except for data used by the contingency SCP, which typically have validity times in the past (they describe unexpected changes, which are typically detected with some delay after they happened).

\begin{figure}
\begin{center}
\includegraphics[width=\hsize]{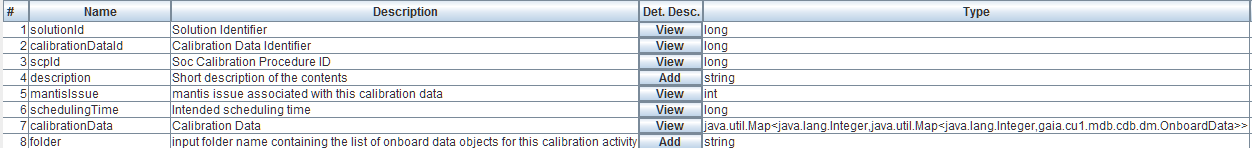}
\end{center}
\caption{Data model for the {\tt CalibrationActivityData} class. The core field of a calibration activity data set is a map of {OnboardData} with two integer keys. The first key corresponds to the group index and the second to the command sequence index. These data sets can unambiguously identified using an unique ID, which is preserved among the different versions, tracked by the solution ID. The SCP ID to which they are associated is also included.}
\label{fig:calibrationActivityData}
\end{figure}

All the on-board data required to run any SCP are stored in objects of class {\tt Calibration\-Activity\-Data} (see Fig.~\ref{fig:calibrationActivityData}. The core field of a calibration activity data set is a map of {OnboardData} with two integer keys. The first key corresponds to the group index and the second to the command sequence index. These data sets can unambiguously be identified by the unique ID. The solution ID keeps track of the version (parser-ingestor execution). The SCP ID to which they are associated is also included. The description is a copy of the SCP description plus some additional useful information. Some metadata required to fully describe the {\bf Uplink} are stored in the CDB as {\tt Uplink} objects. Third, a {\bf Payload Operations Request (POR)} is generated using the Uplink and CDB information. The POR is be stored and sent to MOC.

\begin{figure}
\begin{center}
\includegraphics[scale=0.45]{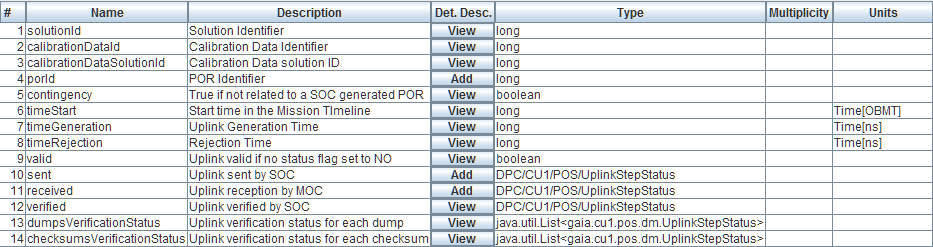}
\end{center}
\caption{Data model for the {\tt Uplink} class. These objects contain all metadata necessary to describe the correct or incorrect execution of a calibration sequence. It can also be used to describe on-board changes not commanded by SOC. It is the key object produced by the scheduler. It includes fields to identify the parent SCP and the child POR. The associated on-board data can be queried in the CDB using the unique solution ID. A flag indicates if the change was commanded by SOC or is used to regularise a different unexpected (anomaly) event. The starting onboard time and generation UTC are also stored. Changes are not retracted, the valid flag and invalid UTC fields are used instead. Up to three flags are used to describe the different steps in the uplink process, which can have the three YES, NO and UNKNOWN values. An Uplink will be valid whenever none of these flags is set to NO.}
\label{fig:uplink}
\end{figure}

The Uplink metadata class structure is shown in Fig.~\ref{fig:uplink}. Each {\tt Uplink} stores a solution ID, the associated SCP and POR identifier, a flag indicating if it was an unintended anomaly change, the start time in the mission timeline, the civil time when it was generated (UTC), status flags with the knowledge available on each uplink step and, finally, the civil time when it was declared invalid, if necessary. All uplinks are considered valid unless they have any step status flag set to NO. That is, the command is assumed to be executed correctly in the absence of confirmation.

\begin{figure}
\begin{center}
\includegraphics[scale=0.45]{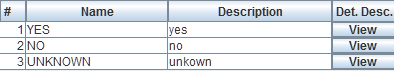}
\end{center}
\caption{Data model for the {\tt UplinkStepStatus} class. It is an enumeration of the knowledge available on the different stages of the uplink process. An uplink will be declared invalid if and only if one or more of the status flags are set to NO. The commands will be assumed correctly executed when no information is available.}
\label{fig:uplinkStepStatus}
\end{figure}

PORs are sent to {\bf MOC}, which provides {\bf MOC data}, including POR reception acknowledgements and spacecraft scientific telemetry and housekeeping data. The consistency between the MOC data and the information in the Uplink and Calibration data bases can be verified in a deterministic way. No invalid {\tt Uplink} or {\tt OnBoardData} object is retracted to solve problems. They are flagged accordingly instead. The contingency tool generates and ingests new anomaly {\tt Uplink} and {\tt OnBoardData}, to make the uplink chain self-consistent again, if required. The preservation of invalid data in the CDB and the inclusion of civil times in the uplink allow the system to reconstruct the CDB values used by DPAC at any time in the past.

Key elements to ensure the consistency of the CDB with the real data on-board Gaia are the VPU memory dumps and checksums, which are downlinked with the spacecraft housekeeping or payload telemetry using the BDT or SIF interfaces. Different types of dumps and checksums have been used during commissioning, each one covering different areas of the VPU parameter blocks. The full area dump is the most widely used, though.

\section{SCP and data set generation language}
\label{sect:scpLanguageParser}

It is basically impossible for a human to self-consistently define the SCP calibration activity templates in a bottom-top approach where the command sequences, command sequence groups, SCPs and data sets are manually generated. A typical SCP contains up to a hundred different command sequences. When the different calibration activities and VPUs are considered, several hundred groups and around a thousand command sequences are required. The data model describing the SCP is also complex, with many ancillary classes and cross-reference indices. This makes the underlying structure barely human-readable, although very efficient and reliable for the POS automatic system.

The strategy adopted has been to define the SCPs as a whole using a descriptive language (pseudo-code) based on the data model. In this way, an SCP is defined by the functionality it requires in a human readable text file. The archives are then parsed by ad-hoc classes which provide the structure required to implement that functionality, in terms of groups and sequences. Ultimately, the command sequences are defined by the ODB command sequence name and formal parameter names and values. The formal parameters are then interpreted by the POS using the default format scheme defined in the ODB. The SCP language parser will be briefly described below.

The text input file is composed of lines. Each line can either contain a single POS language command or blanks. Blank lines are ignored by the parser and can be introduced to increase legibility. Each command line is composed of tokens separated by blanks. Multiple, leading and trailing blanks are ignored, and can be included for the sake of clarity. The first token must always be the command. Commands are always provided in capital letters. The remaining tokens provide attributes, with the format {\tt attributeName=value}. Attribute values including blanks (e.g. descriptions) can be introduced using single quotes ('). Any text within single quotes will not be split in tokens (e.g. {\tt description='Description including blanks'}). The attributes can be provided in any desired order. An attribute cannot be defined twice. Numerical attribute values are interpreted according to Java parsing rules. For example: 0xFF (hexadecimal) and 0377 (octal) are both interpreted as 255 (decimal). Left padding zeroes are thus avoided when providing decimal values, e.g. 08 and 09 are interpreted as invalid octal numbers. Plain decimal notation is the preferred way to specify integers. 

The SCP definition commands are:

  {\bf CALIBRATION\_ACTIVITY} Attribute: calibrationActivity (valid values: String). It defines the calibration
    activity to which this SCP is associated.
    
  {\bf CCD\_ROW} Optional. Attribute: ccdRow (valid values [1,7]). It defines to which VPU 
	  is this SCP associated, if applicable (e.g. contingency is not associated to any VPU).
	  
	{\bf PERMANENT\_CHANGES} Attribute: permanentChanges  (valid values: $\lbrace$true, false$\rbrace$).
	  It defines whether that SCP intends to apply permanent changes on-board. 
	  
	{\bf DESCRIPTION} Attribute: description  (valid values: any string).
	  It defines additional information for the header to be included in the PORs using this SCP.
	  Some default text is automatically generated by the parser-ingestor.
	  
	{\bf GROUP} Attribute: description  (valid values: any string).
	  It provides a descriptive header for a given group of command sequences.
	  All command sequences defined after a group belong to it.
	  At least one group must be created before any command sequence definition.
	  
	{\bf WAIT} Attributes: duration (valid values: any positive integer), unit (valid values:
	  $\lbrace$SECOND, MILLISECOND, NANOSECOND$\rbrace$).
	  It introduces waiting times during the execution of the SCP command sequences.
	  It is useful to allow some systems to settle after a perturbation
	  (table patch, AOCS after go to operational) or as a sort of integration time for some commands
	  such as SIF acquisition. The waiting times are automatically assigned and accumulated
	  to the SCP and group leading offsets or group paddings between sequences.

The command sequences are defined using POS language commands with the same name of the elements in the commandSequenceType POS enum items. The attributes are those fields in the CommandSequence data type directly related to that kind of command sequence. Xstream is used to automatically parse each field to the data type required by the data model. The command sequence POS language commands are:

	{\bf PATCH} Attributes: ccdRow (valid values [1,7]),
	  patchMode (valid values $\lbrace$BDT, SHORT\_LOAD$\rbrace$,
	  patchFull (valid values: $\lbrace$true, false$\rbrace$),
	  onboardDataClass (valid values: any class name extending VpuParameter, e.g.: VoDefinition),
	  patchToDefault (valid values: $\lbrace$true, false$\rbrace$),
	  patchByteOffset (valid values: any non-negative integer, only required if patchFull=false),
	  patchByteLength (valid values: any positive integer, only required if patchFull=false).
	  It patches a VPU table.
	  
	{\bf DUMP} Attributes: ccdRow (valid values [1,7]),
	  dumpChecksumType (valid values: $\lbrace$MOC, SOC, ASTRIUM$\rbrace$,
	  It dumps part of the full VPU tables memory area.
	  
	{\bf CHECKSUM} Attributes: ccdRow (valid values [1,7]),
	  dumpChecksumType (valid values: $\lbrace$MOC, SOC, ASTRIUM$\rbrace$,
	  It computes a checksum of part of the full VPU tables memory area.
	  
	{\bf MODE\_CHANGE} Attributes: ccdRow (valid values [1,7]),
	  modeChangeType (valid values: any string element name in non-data model enum ModeChangeType,
	  e.g: VPU\_OPERATIONAL, SETUP\_IM\-\_LOAD\-\_DEFAULT\_ACPS\-\_NO\_WFS).
	  It tracks Gaia status changes.
	  
	{\bf SIF\_COMMAND} Attributes: ccdRow (valid values [1,7]),
	  sifCommandType (valid values: any string element name in non-data model enum SifCommandType,
	  e.g: OC\_DUMP\_VPU\_SW\_2\_7, RAW\_\-BUFFER\_\-AF8\_AF9WFS\_\-BP\_RP\_\-35612TDI).
	  It defines SIF command sequences.
	  
	{\bf PATCH\_FROM\_VO\_STORE} Attributes: ccdRow (valid values [1,7]),
	  voStorageIndex (valid value: [0-99]),
	  patchToDefault (valid values: $\lbrace$true, false$\rbrace$),
	  It defines a patch to the virtual objects definition VPU table
	  
	{\bf CONTINGENCY} Attributes: none.
	  It defines a way to account for non-SOC scheduled OnboardData changes.
	  A single contingency SCP, group and command sequence shall exist.

An example of valid SCP code is provided below:

\footnotesize
\begin{verbatim}
CALIBRATION_ACTIVITY	calibrationActivity=CHARGE_INJECTION_SCHEME_VPU_SW_2_7				
PERMANENT_CHANGES	peramentChanges=true				
DESCRIPTION	description='POR_IIOC-170_VPU6'				
CCD_ROW	ccdRow=6				

GROUP	description='Declare VPU set unused and switch to service mode'				
  MODE_CHANGE	ccdRow=6	modeChangeType=VPU_UNUSED_SERVICE			

GROUP	description='Set charge injection scheme and charge injection level'				
  PATCH	ccdRow=6	patchMode=BDT	patchFull=true	onboardDataClass=CiParam	patchToDefault=false
  PATCH	ccdRow=6	patchMode=BDT	patchFull=true	onboardDataClass=DefaultCcpTable	patchToDefault=false
  MODE_CHANGE	ccdRow=6	modeChangeType=SETUP_PEM_LOAD_DEFAULT_CCPS			

GROUP	description='Checksum'				
  CHECKSUM	ccdRow=6	dumpChecksumType=ASTRIUM			

GROUP	description='Dump tables for reference'				
  SIF_COMMAND	ccdRow=6	sifCommandType=ASTRIUM_DUMP_VPU_SW_2_7			

GROUP	description='Declare VPU set used'				
  MODE_CHANGE	ccdRow=6	modeChangeType=VPU_USED_OPERATIONAL			

GROUP	description='Dump VPU memory area for MOC'				
  DUMP	ccdRow=6	dumpChecksumType=MOC			
\end{verbatim}
\normalsize

The definition for the contingency SCP, used to track changes not commanded by POS (mostly Airbus commanding, but also anomalies), is also provided below:

\footnotesize
\begin{verbatim}
CALIBRATION_ACTIVITY calibrationActivity=CONTINGENCY
DESCRIPTION description='Contingency SCP'
PERMANENT_CHANGES permanentChanges=true

GROUP description='Contingency group'
  CONTINGENCY
\end{verbatim}
\normalsize

\section{Lessons learned}
\label{sect:lessonsLearned}

During commissioning most of the commanding has been provided by Airbus, which is indicated by the black box in Fig.~\ref{fig:functionalDiagram}. Interleaved Airbus - SOC commanding has been the norm, rather than the exception (as originally expected). This situation has provided very useful interaction, but has put the system under considerable stress, because it was not designed for a short (days, hours), but long term (months) duty cycle.

In particular, the reaction time for the SOC uplink system to put Airbus commanding under configuration control has been reduced down to a few hours. The production to own commanding has also been available in less than a day. These achievements are remarkable when compared to the initial monthly duty cycle. However, Airbus commanding has been very dynamic, with some updates taking place within hours.

The priority from Airbus, MOC and ESA project was to have the best configuration onboard as soon as possible. This approach was convenient to optimise the spacecraft behaviour, but prevented a real-time tracking of the changes onboard. In addition, the DPAC processing was designed as a continuous process without any waiting period (data are processed as soon as they arrive). These two conflicting requirements resulted in some DPAC processing carried out with an inconsistent CDB. The output has sometimes been wrong and retracted from the data base.

DPAC commanding was also needed since the very beginning, contrary to expectations (Airbus commissioning, DPAC nominal phase). However, both the DPAC systems and SOC team were dimensioned for the routine operational phase. This has been a recurrent source of pressure during that time, focused over a small team whose skills were not redundant.

In a mission where correct data processing is so tightly coupled with payload configuration, one should seriously consider the tracking of that payload configuration to be part of the actual pipeline in the form of a pre-processing task.The role of this pre-processing task would be to decode the payload configuration history from the available information in the telemetry stream ,and make that information directly available downstream. This approach would have certainly help in smoothing the real commanding scenario were the latency to provide this configuration as input for the processing was in many cases barely hours.

Several workarounds have been applied to cope with the real commissioning scenario. In some extreme cases, the DPAC pipelines have been paused during several hours (or even 1-2 days) to provide time to track the updates and/or evaluate the system performance when the changes are applied. Additional functionality requirements have also required the DPAC processing to evolve almost continuously, with several software releases taking place during commissioning. It provided the way to cope with new situations. In particular, there was a major update of the on-board software that required a significant rework of the ground segment data processing. This change was carefully planned and all parties (Airbus, MOC, DPAC) were informed well in advance. Finally, several auxiliary systems, not under configuration control, were sometimes required to provide quick reaction (data set or POR generation). The role of these alternative systems has been crucial several times.

\section{Conclusions}
\label{sect:conclusions}

The Gaia commissioning has almost finished. During these six months, both Airbus and DPAC have provided extensive commanding. This situation was unforeseen for a survey mission, but was required to tune the spacecraft behaviour. The DPAC uplink system is composed of the Payload Operations System and the SOC calibration team toolbox and uplink packages. The role of the uplink system is double: on-board commanding and spacecraft status configuration control. They are complemented by several auxiliary modules. Both objectives have been fulfilled, allowing the DPAC pipelines run during this time. The global system architecture and the SCP language parser have been described in detail.

Problems have also been experienced and discussed. Most of them were related to the DPAC system been prepared for a passive role steady state situation, and not a very dynamic environment with multiple changes per week and substantial DPAC commanding required. All problems have been solved, and none has been a show stopper for DPAC processing. An account of the lessons learned has also been provided. They could be particularly interesting for future space survey missions.

\section{Acknowledgements}

The authors wish to thank the Gaia MOC operations and Airbus Defence \& Space teams for their continuous support and access to internal documents. Some concepts and ideas presented here come from those sources.

Material used in this work has been provided by Coordination Unit 1 (CU1) of the Gaia Data Processing and Analysis Consortium (DPAC). It is gratefully acknowledged for their contribution.

\bibliography{2014_06_spie_uplink,gaia_livelink_valid,gaia_livelink_obsolete,gaia_drafts,gaia_refs,gaia_books,gaia_refs_ads}   
\bibliographystyle{spiebib}   

\end{document}